\documentclass{IEEEtran4PSCC}
\ifCLASSINFOpdf
   \usepackage[pdftex]{graphicx}
\else
   \usepackage[dvips]{graphicx}
\fi
\usepackage{dsfont}

\usepackage[cmex10]{amsmath}
\usepackage{amsfonts}
\usepackage{algorithm}
\usepackage{algpseudocode}

\usepackage{ntheorem}
\theoremseparator{:}

\usepackage{fnpct}
\ifCLASSOPTIONcompsoc \usepackage[caption=false,font=normalsize,labelfont=sf,textfont=sf]{subfig}
\else
\usepackage[caption=false,font=footnotesize]{subfig}
\fi
\usepackage{tikz}
\makeatletter
\let\old@ps@headings\ps@headings
\let\old@ps@IEEEtitlepagestyle\ps@IEEEtitlepagestyle
\def\psccfooter#1{%
    \def\ps@headings{%
        \old@ps@headings%
        \def\@oddfoot{\strut\hfill#1\hfill\strut}%
        \def\@evenfoot{\strut\hfill#1\hfill\strut}%
    }%
    \def\ps@IEEEtitlepagestyle{%
        \old@ps@IEEEtitlepagestyle%
        \def\@oddfoot{\strut\hfill#1\hfill\strut}%
        \def\@evenfoot{\strut\hfill#1\hfill\strut}%
    }%
    \ps@headings%
}
\makeatother

\begin{document}
\title{Coordinated Day-ahead Dispatch of Multiple Power Distribution Grids hosting Stochastic Resources: An ADMM-based Framework}
\author{
    \IEEEauthorblockN{
    Rahul Gupta, Sherif Fahmy, Mario Paolone\\}
    \IEEEauthorblockA{Distributed Electrical Systems Laboratory, EPFL, Lausanne, Switzerland\\
    \{rahul.gupta, sherif.fahmy, mario.paolone\}@epfl.ch
    \\}
    }

% make the title area
\maketitle
\begin{abstract}
This work presents a framework to compute the aggregated day-ahead dispatch plans of multiple and interconnected distribution grids operating at different voltage levels. Specifically, the proposed framework optimizes the dispatch plan of an upstream medium voltage (MV) grid accounting for the flexibility offered by downstream low voltage (LV) grids and the knowledge of the uncertainties of the stochastic resources.
The framework considers grid, i.e., operational limits on the nodal voltages, lines, and transformer capacity using a linearized grid model, and controllable resources' constraints. The problem is formulated as a stochastic-optimization scheme considering uncertainty on stochastic power generation and demands and the voltage imposed by the upstream grid.
The problem is solved by a distributed optimization method relying on Alternating Direction Method of Multipliers (ADMM) that splits the main problem into one aggregator problem solved at the MV-grid level and several local problems solved at the MV-connected-controllable-resources and LV-grid levels.
The use of distributed optimization enables privacy-aware dispatch computation where the centralized aggregator is agnostic of the parameters of the participating resources and downstream grids.
The framework is validated for interconnected CIGRE medium- and low-voltage networks hosting heterogeneous stochastic and controllable resources.
\end{abstract}

\begin{IEEEkeywords} 
Dispatching, multi-grid dispatch, distributed optimization, stochastic optimization, privacy-aware, ADMM.
\end{IEEEkeywords}

\section{Introduction}
Dispatching power distribution networks to track a predefined dispatch plan at the grid connection point (GCP) has been proposed in the literature as a way to achieve the coordination between transmission and distribution networks with the aim of reducing the activation of expensive reserves \cite{chowdhury1990review}. Conventionally, distribution system operators (DSOs) compute their dispatch plan the day before its operation after it has been cleared by the electricity market \cite{happ1974optimal, lee1985united}. 
However, the increasing connection of stochastic renewables (e.g. Photovoltaic (PV) plants, wind farms and small hydro units) and loads (e.g. electric-vehicles), makes it difficult to predict and follow the day-ahead-dispatch-plan with high fidelity. 
In the recent literature (e.g., \cite{sossan2016achieving}), the dispatch of distribution networks by means of controllable resources such as utility-scale battery energy storage system (BESS) has been proposed and experimentally validated. 
In \cite{gupta2020grid} a local microgrid was dispatched with the help of controllable BESS and two curtailable PV plants. Both methods proposed in these papers rely on a two-stage framework. The first stage is the scheduling phase, where a dispatch plan is computed knowing the flexibility offered by the controllable resources and the forecasting uncertainty of the prosumption. The second stage is the real-time control phase, where the computed dispatch plan is tracked by actuating controllable resources (e.g., BESS).
The work in \cite{sossan2016achieving} was developed for a medium voltage (MV) feeder, whereas \cite{gupta2020grid} considered dispatching a low voltage (LV) grid. Although the proposed frameworks successfully achieve MV- and LV-grid dispatchability, they assume them to be separate standalone systems. % when they are not connected directly to each other. 
As a matter of fact, in the case of interconnected systems, the fluctuating intra-grid power-flows 
should not be neglected, e.g. downstream networks can affect each other. 
%in the downstream network will affect the other. 
The conventional way to model downstream networks when controlling upstream grids is to forecast their aggregate power-flows. However, this might not be the optimal choice if the former can provide flexibility to the latter (and vice-versa).
Note that, even though it is more optimal to consider downstream-grids when controlling upstream ones, accounting for the flexibility and uncertainty of the downstream systems into the dispatch problem of the upstream system may rapidly increase the number of decision variables. Also, the network parameters of all networks are not publicly available and not shared. 

Some existing works have addressed the multi-grid dispatch problem. For example, in \cite{nikmehr2015optimal} authors propose a multi-microgrid optimal dispatch algorithm where stochastic and probabilistic models of microgrids and energy resources are embedded into the day-ahead-dispatch-plan computation. The problem is solved using a heuristic-based algorithm, and does not include the operational constraints of the grid. In \cite{contreras2018improved}, a method to evaluate the flexibility of the distribution grids that can be provided to the upstream MV system, is proposed. More specifically, the problem first quantifies the flexibility offered by different units within the downstream grid, then aggregates and send the result of the first-phase to the MV system. 
However, the computation is unidirectional and, therefore, does not consider other sources of flexibility (e.g. other resources). 
In \cite{kahl2014cooperative} a cooperative dispatch scheme is proposed where the transmission network was divided into different areas where their respective dispatch-plans were computed by using a distributed optimization framework. Grid constraints are also ignored in this approach. Finally, the work in \cite{fortenbacher2014grid} solves a multi-level power system dispatch problem where the grid constraints are modeled by leveraging the DC-approximation of the load-flow equations. However, as known, the DC-flow approximation of the power-flow equations does not yield accurate results for distribution grids.
%However, the grid constraints consider a dc-grid model, which might not be feasible for realistic distribution grids.

Following the shortcomings of the above-mentioned works, this paper proposes an optimization framework capable to determine the aggregated dispatch of a MV network considering the flexibility of multiple downstream LV networks and controllable resources (e.g. BESSs). A direct consequence of the proposed formulation is also the determination of the aggregate dispatch-plans of the different MV-connected LV-downstream-grids. The problem is formulated using an Alternating Direction Method of Multipliers (ADMM)-based distributed optimization scheme that guarantees a inter-grid-layer-privacy. It consists of a main aggregator and sub-problem-solvers. The aggregator is the MV network DSO, whereas the sub-problem-solvers are MV-connected controllable entities 
%the local objectives of the participating resources 
such as resources (e.g. BESS) and downstream LV systems. The aggregator solves an optimal power flow (OPF) considering the uncertainty of both the upper-layer-transmission-grid nodal voltages and MV-connected loads/generating units. The connected downstream LV systems also solve an OPF that accounts for the uncertainty of both nodal voltages imposed at the MV/LV interfaces and the LV-connected loads/generating units. Other MV-connected-controllable-distributed-resources also solve local optimization problems maximizing various utility functions while accounting for local constraints (e.g. a BESS ensuring its capacity constraints while maximising usage profit). 
The aggregator and the local subproblems are derived and solved using the ADMM decomposition \cite{boyd2011distributed}. 
%This structure was first used in author's previous work \cite{gupta2020grid} for real-time distributed control of the resources. 
In short, the framework is formulated as a privacy-aware distributed optimization scheme where participating resources do not share their models.
It therefore enables a distributed privacy-aware computation of the day-ahead dispatch plan accounting for the flexibility provided by several downstream networks and MV-connected-controllable-resources (e.g. BESS and PV units) without the explicit knowledge of their network or resource models, respectively. 
%Furthermore, the problem formulation also considers the flexibility supplied by distributed resources such as BESS and PV units. The problem is solved by distributed optimization using the Alternating Direction Method of Multipliers (ADMM), where the centralized dispatch problem into decomposed into local problems connected with Lagrange multipliers. The problem is solved iteratively until it reaches convergence.
%The grid constraints are modeled using sensitivity-coefficients based linear constraints on nodal voltages, lines ampacities, and losses.
Additionally, to address the shortcomings of the works in the literature, grid-constraints are included in both the main aggregator and LV-subproblem OPFs by leveraging sensitivity coefficients-based linearizations of the power-flow equation. As a result, grid-losses are accounted for and linear constraints on nodal voltage and branch current magnitudes are included, in all the proposed OPFs.
Finally, the proposed problem is validated through simulations performed on a MV system connected with two LV systems, one MW-scale battery, and several PV generation units. %We also compare the solution obtained by the ADMM based distributed optimization with centralised solution.

The paper is organized as follows. Sec. II defines the problem statement, Sec. III presents the multi-grid dispatch formulation and its decomposition into distributed problems. Sec. IV presents the test cases and results, and finally, Sec. V concludes the work.

\section{Problem Statement}
\begin{figure}[h]
    \centering
    \includegraphics[width=0.95\linewidth]{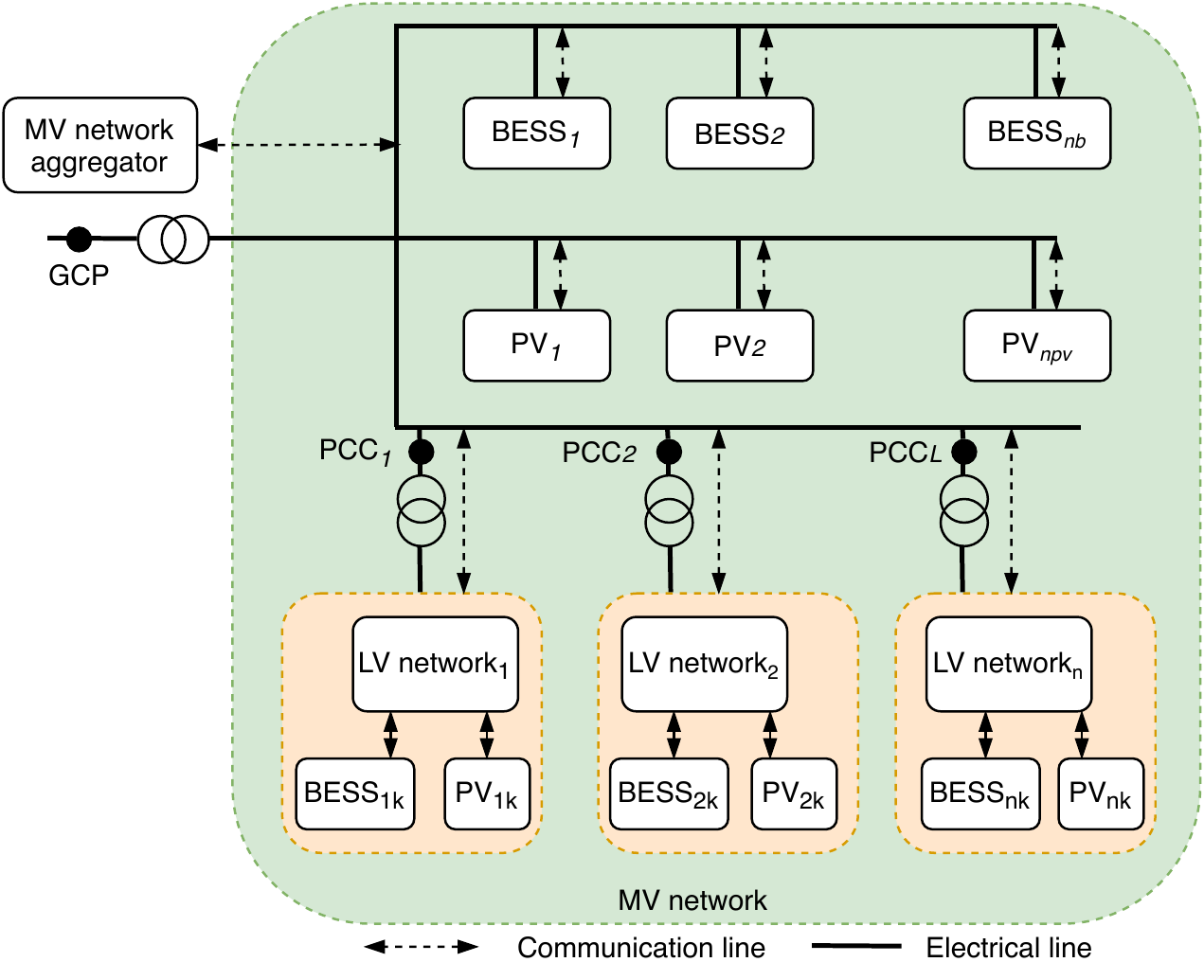}
    \caption{Schematic showing MV network architecture connected with multiple controllable and stochastic resources and downstream low voltage networks. The solid and dotted lines represent electrical and communication connections respectively.}
    \label{fig:architecture}
\end{figure}

Let us consider a MV distribution grid interfaced with multiple stochastic and controllable resources which can be controlled in real-time (e.g., PVs or BESSs).  The MV grid is also interfaced with downstream LV distribution grids through MV/LV transformers. The LV systems also host stochastic resources such as PV plants and flexible resources such as BESSs. 
Figure~\ref{fig:architecture} shows the scheme of the considered power-system setup where the solid and dotted lines represent electrical and communication connections, respectively.
The main objective of the proposed framework is to compute an optimal day-ahead dispatch at the grid connection point (GCP) of the MV grid. The day-ahead dispatch plan should account for
\begin{itemize}
    \item the stochasticity of the electricity demand and generation in both MV and the LV networks;
    \item the flexibility and capability limits of the controllable resources in both MV and the LV networks; 
    \item the operational constraints of MV and LV grids i.e. nodal voltage magnitudes within bounds, branch ampacities; 
    %\item the capability limits of the controllable resources.
\end{itemize}
To mimic real-world commercial privacy, we assume that the parameters of the LV networks and the MV-connected-controllable-resources (MVCCRs), willing to participate in the dispatching, are not known to the MV aggregator. However, the former can share relevant information for the GCP dispatch plan to be optimally computed. More specifically, each LV system shares the power-flows and nodal voltage magnitudes at its point of common coupling (PCC) (see Fig.~\ref{fig:architecture}), and each MVCCR shares the apparent power flexibility they can offer within their capabilities.

%It is a relevant problem to solve as the MV systems are often connected to several LV systems and the controllable resources for which the models can not be shared explicitly. The methods proposed in the literature \cite{fortenbacher2014grid} require knowledge of the models of the participating resources, which might not be realistic. In this context, we solve this problem by a privacy-aware distributed optimization scheme where the centralized problem is decomposed into individual resources problems coordinated by a central aggregator. The resource problems solve their local subproblems without the need to share their system models. We use ADMM-based distributed optimization to decompose the problem into resource-specific problems, which are solved iteratively until convergence. 

%In both the grids, the grid constraints
In\footnote{In this work vectors and matrices are denoted by bold symbols. Complex number are denoted with a bar (e.g. $\bar{z} = |\bar{z}|\exp(j\theta)$) while their complex conjugates are underlined (e.g. $\underbar{z}$)} all OPFs of this work, the grid operational constraints are modeled using sensitivity coefficient-based linearized power flow model (e.g.\cite{christakou2013efficient}). Let\footnote{All non-slack nodes are modelled as PQ-injection nodes.} $N_b$ be the number of non-slack buses in the considered MV grid, $N_l$ the number of lines, $\mathbf{\bar{V}} \in \mathds{C}^{N_b}$ the nodal voltages, $\mathbf{\bar{I}} \in \mathds{C}^{N_l}$ the branch currents, $\mathbf{P}, \mathbf{Q} \in \mathds{R}$ the vectors containing controllable active and reactive injections, $P_d, Q_d \in \mathds{R}$ the active and reactive aggregate grid losses; the grid model can be expressed for time $t$ as
\begin{align}
|\mathbf{\bar{V}}_t|&=\mathbf{A}_t^{v} \begin{bmatrix}
    \mathbf{P}_t\\
    \mathbf{Q}_t
\end{bmatrix}
+ \mathbf{b}_t^{v}\label{eq:lin_grid_model_v}\\
|\mathbf{\bar{I}}_t|&=\mathbf{A}_t^{{i}} \begin{bmatrix}
    \mathbf{P}_t\\
    \mathbf{Q}_t
\end{bmatrix}
+ \mathbf{b}_t^{i}\label{eq:lin_grid_model_i}\\
\begin{bmatrix}
    P_{d,t}\\
    Q_{d,t}
\end{bmatrix} &=\mathbf{A}_t^d \begin{bmatrix}
    \mathbf{P}_t\\
    \mathbf{Q}_t
\end{bmatrix} 
+ \mathbf{b}_t^{d},
\label{eq:lin_grid_model_pl}
\end{align}
%\end{subequations}
where $ \mathbf{P}_t \in \mathds{R}^{N_b}$ and  $\mathbf{Q}_t \in \mathds{R}^{N_b}$ are, respectively, nodal active and reactive power injections, $\mathbf{A}_t^{v} \in \mathds{R}^{N_b\times 2N_b}$, $\mathbf{A}_t^{{i}} \in \mathds{R}^{N_l\times 2N_b}$, $\mathbf{A}_t^{d} \in \mathds{R}^{2 \times 2 N_b}$, $\mathbf{b}_t^{v} \in \mathds{R}^{N_b} $, $\mathbf{b}_t^{{i}}  \in \mathds{R}^{N_l}$, and $\mathbf{b}_t^{d} \in \mathds{R}^2$ are state-dependent parameters enabling power-flow-linearizations, composed of constants and sensitivity coefficients (SCs) for time index $t$. SCs are formally defined as the partial derivative of controllable electric quantities (e.g. branch currents) w.r.t. control variables (i.e. nodal power injections).
They are determined with the method in \cite{christakou2013efficient} by solving a system of linear equations (that admits a single solution, as proven in \cite{paolone2015optimal}) as a function of the grid's admittance matrix and the system state (i.e. nodal voltages at all buses). In the day-ahead phase, the SCs along the whole scheduling horizon are pre-computed using point predictions of the nodal injections. In the real-time phase, the SCs are updated at each control step using the present grid state.
In the following, the linearization parameters for MV and $l$-th LV systems for time $t$ are denoted by $\mathbf{A}^{\text{mv},v}_t, \mathbf{A}_t^{\text{mv},i}, \mathbf{A}_t^{\text{mv},d},\mathbf{b}_t^{\text{mv},v}, \mathbf{b}_t^{\text{mv},i}, \mathbf{b}^{\text{mv},d}_t$ and $\mathbf{A}_t^{\text{lv}_l,v},$ $\mathbf{A}_t^{\text{lv}_l,i},$ $\mathbf{A}_t^{\text{lv}_l,d}$, $\mathbf{b}_t^{\text{lv}_l,v}, \mathbf{b}_t^{\text{lv}_l,i}, \mathbf{b}_t^{\text{lv}_l,d}$ respectively.
%

% DONE TILL HERE %

%%%%%%%%%%%%%%%%%%%%%%%%%%%%%%%%%%%%%%%%%%%%%%%%%%%%%%%%%%%%%%
%%%%%%%%%%%%%%%%% Day-ahead dispatch %%%%%%%%%%%%%%%%%%%%%%%%%
%%%%%%%%%%%%%%%%%%%%%%%%%%%%%%%%%%%%%%%%%%%%%%%%%%%%%%%%%%%%%%
\section{Multi-grid Day-ahead Dispatch Problem} \label{sec:Day_ahead}
In\footnote{In the following all electric quantities are expressed in per unit as we are dealing with multi-voltage-layer grids.} the following, we present the aggregated day-ahead dispatch problem -- referred to as multi-grid day-ahead dispatch -- for a multi-grid system, i.e. MV grid connected to several LV networks and MVCCRs. 
%for multi-grid system (referred to as multi-grid day-ahead dispatch) i.e., for the MV system connected with several LV grids and controllable resources such as BESS. 
First, the centralised problem is presented, then, to guarantee privacy and mimic real-world intra-grid-operator-relationships, i.e. the non-availability of LV downstream grid and MVCCR models at the MV aggregator level, a decomposition, leveraging distributed optimization relying on the ADMM, of the centralised dispatch problem is presented. 
%but, since the models of the LV grid and controllable resources are not known to the MV aggregator, we decompose the dispatch problem using privacy-aware distributed optimization using ADMM.
%
\subsection{Design requirements of the dispatch plan}
%The objective is compute aggregated day-ahead dispatch plan of the MV system connected with multiple LV systems and controllable resoruces such as BESS. 
The main objective is to compute a day-ahead dispatch plan, i.e. the active power trajectory, that the MV distribution network advertises to its upstream network and should follow at its GCP during the next day of operation. As previously mentioned, since downstream LV network uncertainties and flexibilities are accounted for in the latter problem, the day-ahead dispatch plans at the different PCCs of the downstream grids are byproducts of the problem resolution.  
%The aggregated dispatch plan accounts for the uncertainty and flexibility of the downstream LV systems. 
The design requirements of the proposed dispatch plan are: 
\begin{itemize}
    \item Stochastic variations from distributed generation and demand should be compensated by the LV-networks and MVCCRs while respecting their operational constraints.;
    \item Voltage uncertainties at the MV-GCP and, as a result, at the different LV-PCCs should be accounted for;
    \item The regulation made by MVCCRs and the controllable resources in the LV-downstream-grids should not violate MV and LV grid operation constraints.
\end{itemize}
The dispatch plan is computed with a stochastic optimization framework, where the stochasticity of distributed generation and demand nodal power injections is captured through scenarios. As previously discussed, grid operational constraints are modelled with a linearized grid model. Operational constraints of the controllable resources are modelled accounting for the PQ-capability-sets of their power converters and, in case these resources are BESS, also by state-of-energy constraints. %and for energy-constrained resources as batteries, as discussed in the Section~\ref{sec:Expt.}.
\subsection{Centralized problem formulation}
Let $\mathcal{L}:= l = 1, \dots, L$ and $\mathcal{R}:= r = 1, \dots, R$ be the set of indices of LV grids and MVCCRs, respectively that are connected to MV network. Let $\mathcal{T} = [t_0, t_1 \dots, t_{N}]$ be the set of time indices of the scheduling horizon delimited by $t_0$ and $t_N$. We assume the set $\Omega = \Omega^{\text{mv}} \cup \bigcup\limits_{l=1}^{\mathcal{L}} \Omega^{\text{lv}_l}$ collects the scenarios $\omega$ for stochastic nodal-power-injections of uncontrollable generation and demand units, where $\Omega^{\text{mv}}$ and $\Omega^{\text{lv}_l}$ are scenario sets for the MV and $l-$th LV networks, respectively.
$\mathbf{P}_t^{\omega},\mathbf{Q}_t^{\omega}$ and $\mathbf{P}_{t, \text{unc}}^{\omega},\mathbf{Q}_{t, \text{unc}}^{\omega}$ contain controllable, i.e. aggregate LV-PCC and MVCCRs, and uncontrollable nodal active/reactive power injections of the MV system for timestep $t$ and scenario $\omega$. 
Similarly, $\mathbf{p}_t^{l,\omega},\mathbf{q}_t^{l,\omega}$ and $\mathbf{p}_{t, \text{unc}}^{l,\omega},\mathbf{q}_{t, \text{unc}}^{l,\omega}$ contain the nodal active/reactive controllable and uncontrollable power injections for the $l-$ LV system ($l\in\mathcal{L}$).
Let $\bar{S}_{0,t}^{\omega} = {P}_{0,t}^{\omega} + j{Q}_{0,t}^{\omega}$ be the nodal-apparent-power-injections at the slack bus of the MV network (i.e. GCP), $\bar{s}_{0,t}^{l,\omega} = p_{0,t}^{l,\omega} + jq_{0,t}^{l,\omega} = P_{l,t}^{\omega} + jQ_{l,t}^{\omega}$ the nodal-apparent-power-injections at slack bus of $l-$th LV network ($l-$th PCC), $\bar{v}_{0,t}^{l,\omega} = \bar{V}_{l,t}^{\omega}$ the nodal voltage at the $l$-th PCC, $\bar{S}^{\text{disp}}_t = P^{\text{disp}}_t + j Q^{\text{disp}}_t$ the decision variable for the main MV-dispatch plan for time $t$, where $P^{\text{disp}}_t$ and $Q^{\text{disp}}_t$ refer to the active and reactive nodal powers, $\bar{s}^{l,\text{pcc}}_t = p^{l,\text{pcc}}_t + j q^{l,\text{pcc}}_t$ the auxiliary decision variable for the $l-$th LV network-dispatch plan at the $l$-th PCC. The variables $P_{r,t}^{\omega}, Q_{r,t}^{\omega}$ denote the active and reactive nodal power injections of MVCCRs $r\in\mathcal{R}$. The $l-$th LV network is also connected with controllable resources with indices defined by set $\mathcal{R}_l:=r_l = 1,\dots, R_l$ with active and reactive nodal power injections (decision variables) denoted by $p_{r,t}^{l,\omega}, q_{r,t}^{l,\omega}$ for time $t$ and scenario $\omega$. The symbols $p_{d,t}^{l,\omega}, q_{d,t}^{l,\omega}$ denote aggregated active and reactive grid losses of $l-th$ LV network for time $t$ and scenario $\omega$. 

The main idea behind the proposed formulation is to determine a main dispatch plan at the MV network GCP such that it can be tracked for any of the forecasted scenarios. The problem consists in determining the injections of the controllable resources (in both MV and LV grids) so as to minimize:
\begin{itemize}
    \item The deviation between the MV dispatch plan $\bar{S}^{\text{disp}}$ and its slack (i.e. GCP) nodal apparent power injections $S_{0}^{\omega}$ for all the scenarios  $\omega \in \Omega$ and timesteps $t \in \mathcal{T}$;
    \item The deviation between the LV dispatch plan $\bar{s}^{l,\text{pcc}}$ and its slack (i.e. $l-th$ PCC) nodal apparent power injections $\bar{s}_{0}^{l,\omega}$ for all the scenarios $\omega \in \Omega$, timesteps $t \in \mathcal{T}$ and all LV-grids $l \in \mathcal{L}$;
    \item MV-resource-specific costs $f_r(P_{r,t}, Q_{r,t})$ that reflect the willingness of each MVCCR to provide regulating power (specific cost functions can be found in section \ref{sec:BESS_cost});
    \item LV-controllable-resource-specific costs $f^l_{r}(p^l_{r,t}, q^l_{r,t})$ that reflect their willingness to provide regulating power (specific cost functions can be found in section \ref{sec:BESS_cost})
\end{itemize}
%{
%Moreover, the cost function includes a resource-specific term  and a coefficient $\lambda_r$ to weight them. Both the cost function and the coefficient can be designed by the modeller, for instance, based on a combination of resource’s operating conditions (such as minimize wear and tear, power ramping, power variations etc.) or the monetary cost associated to it's operation. The cost function should be convex in order to keep the convexity of the overall problem formulation. 
%for instance, on ageing considerations. 
%The impact of the term $\lambda_r$ is investigated in the Appendix.
%}
The proposed centralised problem can therefore be written as\footnote{The symbol $\left\|.\right\|$ refers to norm-2.}:
\small
\label{eq:day_ahead_form}
\begin{align}
\begin{aligned}
    \hat{\bar{S}}^{\text{disp}} = \underset{\begin{matrix}
                \scriptstyle \bar{s}^{l,\text{pcc}}, \bar{S}^{\text{disp},}\\
                \end{matrix}}
    {\text{arg min}} & \sum_{\omega \in \Omega}\sum_{t \in \mathcal{T}} \Bigg\{\left\| \bar{S}_{0,t}^{\omega} - \bar{S}^{\text{disp}}_t \right\|^2 + \sum_{r\in\mathcal{R}} f_r(P_{r,t}^{\omega}, Q_{r,t}^{\omega})\\ +
    \sum_{l\in\mathcal{L}}\Big(& \left\| \bar{s}_{0,t}^{l,\omega} - s^{l,\text{pcc}}_t \right\|^2 + \sum_{r\in\mathcal{R}_l} f^l_r(p_{r,t}^{l,\omega}, q_{r,t}^{l,\omega})\Big)\Bigg\} \label{eq:costdispatch}
\end{aligned}
\end{align}
\normalsize
subject to the following constraints.
\subsubsection{Constraints of the MV system}
\begin{subequations}
\label{eq:MVconst}
the linearised power flow at the GCP and grid losses equality constraints. The linearizations w.r.t. nodal power injections are given by \eqref{eq:tracking_error_pdis_MV} - \eqref{eq:loss_model_disp_MV}.
%of MV system as a function of the nodal injections and the losses
%\footnotesize
%\small
\begin{align}
&  {P}_{0,t}^{\omega}  = \sum_{r\in\mathcal{R}} P_{r,t}^{\omega} + \sum_{l\in\mathcal{L}} P_{l,t}^{\omega} + \mathbf{1}^T\mathbf{P}_{t,\text{unc}}^{\omega} +  P^{\omega}_{d,t}  ~\forall t \in \mathcal{T}, \omega \in \Omega \label{eq:tracking_error_pdis_MV},\\
&  {Q}_{0,t}^{\omega} = \sum_{r\in\mathcal{R}} Q_{r,t}^{\omega} + \sum_{l\in\mathcal{L}} Q_{l,t}^{\omega} + \mathbf{1}^T\mathbf{Q}_{t,\text{unc}}^{\omega}  + Q^{\omega}_{d,t}  ~\forall t \in \mathcal{T}, \omega \in \Omega \label{eq:tracking_error_qdis_MV}, \\
& \begin{bmatrix}
P^{\omega}_{d,t}\\
Q^{\omega}_{d,t}
\end{bmatrix}  = 
\mathbf{A}^{\text{mv},d}_{t, \omega}
\begin{bmatrix}
    \mathbf{P}^{\omega}_t\\
    \mathbf{Q}^{\omega}_t
\end{bmatrix} +
\mathbf{{b}}^{\text{mv},d}_{t, \omega} ~\forall t \in \mathcal{T}, \omega \in \Omega \label{eq:loss_model_disp_MV},
\end{align}
\normalsize
The minimum power factor constraint\footnote{The constraints is non-convex and infeasible when the real power at the GCP is zero. We use the convexification approach proposed in \cite{stai2017dispatching}. They are briefly described in Appendix~\ref{sec:convexification}.} at the GCP imposed by a $\cos(\theta)_{\text{min}}$
\begin{align}
& {|{P}_{0,t}^{\omega}|}/{|{\bar{S}}_{0,t}^{\omega}|} \geq  {\cos(\theta)_{\text{min}}} & ~\forall t \in \mathcal{T}, \omega \in \Omega \label{eq:Q_cons_disp},
\end{align}
The limits on the nodal voltages by bound magnitudes $[V^{\text{min}},V^{\text{max}}]$ and currents $\mathbf{I}^{\text{max}}$ by lines' ampacities)
\begin{align}
& V^{\text{min}} \leq \mathbf{A}^{\text{mv},{v}}_{t, \omega} \begin{bmatrix}
    \mathbf{P}^{\omega}_t\\
    \mathbf{Q}^{\omega}_t
\end{bmatrix}
+ \mathbf{b}^{\text{mv}, v}_{t,\omega} \leq V^{\text{max}}  ~\forall t \in \mathcal{T}, \omega \in \Omega \label{eq:volt_const_disp},\\
    & 0 \leq \mathbf{A}^{\text{mv},{i}}_{t, \omega} \begin{bmatrix}
    \mathbf{P}^{\omega}_t\\
    \mathbf{Q}^{\omega}_t
\end{bmatrix}
+ \mathbf{b}^{\text{mv},{i}}_{t, \omega} \leq \mathbf{I}^{\text{max}}  ~~~\forall t \in \mathcal{T}, \omega \in \Omega \label{eq:current_const_disp}.
\end{align}
\end{subequations}
\subsubsection{Constraints for MVCCRs} PQ-capability-set limits denoted by\footnote{The resource objectives and constrained are detailed in Sec.\ref{sec:BESS_cost}.}
\begin{align}
\Phi_r(P_{r,t}^{\omega},Q_{r,t}^{\omega}) & \leq 0 ~~\forall t \in \mathcal{T}, \omega \in \Omega, r \in\mathcal{R} \label{eq:Resconst}.
\end{align}
\subsubsection{Constraints of the LV systems $\forall l\in\mathcal{L}$} the linearized power flows at the $l$-th PCC and grid losses equality constraints. The linearizations w.r.t. nodal power injections are given by \eqref{eq:tracking_error_pdis} - \eqref{eq:loss_model_disp}.
\begin{subequations}
\label{eq:LVconst}
\begin{align}
{p}_{0,t}^{l,\omega}  & = \sum_{r\in \mathcal{R}_l} p_{r,t}^{l,\omega} + \mathbf{1}^T\mathbf{p}_{t,\text{unc}}^{l,\omega} +  p^{l, \omega}_{d,t}  ~\forall t \in \mathcal{T}, \omega \in \Omega \label{eq:tracking_error_pdis},\\
 {q}_{0,t}^{l,\omega} & = \sum_{r\in \mathcal{R}_l}  q_{r,t}^{l,\omega} + \mathbf{1}^T\mathbf{q}_{t,\text{unc}}^{l,\omega} + q^{l, \omega}_{d,t}  ~\forall t \in \mathcal{T}, \omega \in \Omega \label{eq:tracking_error_qdis}, \\
 \begin{bmatrix}
p^{l, \omega}_{d,t}\\
q^{l, \omega}_{d,t}
\end{bmatrix} &  = 
\mathbf{A}^{\text{lv}_l,d}_{t, \omega}
\begin{bmatrix}
    \mathbf{p}^{l,\omega}_t\\
    \mathbf{q}^{l,\omega}_t
\end{bmatrix} +
\mathbf{b}^{\text{lv}_l,d}_{t, \omega} ~\forall t \in \mathcal{T}, \omega \in \Omega \label{eq:loss_model_disp}.
\end{align}
The voltage imposed by the MV system at LV's PCC is 
\begin{align}
 |\bar{V}^{\omega}_{l,t}| & = |\bar{v}_{0,t}^{l,\omega}| ~\forall t\in\mathcal{T}, \omega \in \Omega, \label{eq:MV_LV_Vcouple}
\end{align}
The minimum power factor constraint at the $l$-th PCC imposed by $\cos(\theta)_{\text{min},l}$
\begin{align}
& {|{p}_{0,t}^{l,\omega}|}/{|{\bar{s}}_{0,t}^{l,\omega}|} \geq  {\cos(\theta)_{\text{min},l}} & ~\forall t \in \mathcal{T}, \omega \in \Omega \label{eq:Q_cons_disp_LV}.
\end{align}
The nodal voltages magnitudes bounded by voltage limits $[v^{\text{min},l}, v^{\text{max},l}]$ and branch current magnitude constraints by its ampacities $\mathbf{i}_l^{\text{max}}$ given by
\begin{align}
& v^{\text{min},l} \leq \mathbf{A}^{\text{lv}_l,v}_{t,\omega} \begin{bmatrix}
    \mathbf{p}^{\omega}_t~
    \mathbf{q}^{\omega}_t
\end{bmatrix}^T 
+ \mathbf{b}^{\text{lv}_l,l}_{t, \omega} \leq v^{\text{max},l}  ~\forall t \in \mathcal{T}, \omega \in \Omega \label{eq:volt_const_disp},\\
    & 0 \leq \mathbf{A}^{\text{lv}_l,i}_{t, \omega} \begin{bmatrix}
    \mathbf{p}^{\omega}_t~
    \mathbf{q}^{\omega}_t
\end{bmatrix}^T 
+ \mathbf{b}^{\text{lv}_l,i}_{t, \omega} \leq \mathbf{i}_l^{\text{max}}  ~~~\forall t \in \mathcal{T}, \omega \in \Omega \label{eq:current_const_disp}.
\end{align}
\end{subequations}
The constraints of the controllable resources\footnote{It is assumed that LV networks know the models of their controllable resources; thus, they are included within LV constraints. This is unlike MVCCRs, of which models are not known to the MV network.} connected to LV networks are denoted by 
\begin{align}
\Phi^l_r(p_{r,t}^{l,\omega},q_{r,t}^{l,\omega}) & \leq 0 ~~\forall t \in \mathcal{T}, \omega \in \Omega, r\in\mathcal{R}_l\label{eq:ResconstLV}.
\end{align}

Once the problem in \eqref{eq:day_ahead_form} is solved, the dispatch plans are the real part of its solution $\hat{S}^{\text{disp}}, \hat{s}^{l,\text{pcc}}$:
\begin{align}
& \hat{P}^{\text{disp}} = \Re\left\{\hat{\bar{S}}^{\text{disp}}\right\}, 
& \hat{p}^{l,\text{pcc}} = \Re\left\{\hat{\bar{s}}^{l,\text{pcc}}\right\} && \forall l \in \mathcal{L}. \label{eq:dispatchplan}
\end{align}

%\subsection{BESS model}
\subsection{ADMM-based privacy-aware decomposition}
\label{sec:ADMM_dispatch}
%%%%%%%%%%%%%%%%%%%%%%%%%%%%%%%%%%%%
%--------Privacy-aware decomposition---
%%%%%%%%%%%%%%%%%%%%%%%%%%%%%%%%%%%%%
The problem formulation in Sec~\ref{eq:day_ahead_form} is centralized, thus, it requires to know the parameters and model of each of the connected LV network and resources. However, this is not practical in real-life. 
Also, it has poor scalability due to increased amount of decision variables. 
Using ADMM-based distributed optimization, this problem can be reformulated into individual local problems and one global problem which can be solve iteratively.
The problem is a standard sharing problem and separable in original decision variables of the local problems. It can be solved in a distributed manner by each resources; then, the solutions from each local problem are sent to the aggregator that accounts for the global constraints and objectives. 
It is achieved by introducing set of auxiliary variables into the aggregator problem which mimic the solutions of the local subproblems. We introduce $\tilde{P}^{\omega}_{r,t}, \tilde{Q}^{\omega}_{r,t}$ that represent local variables for active and reactive powers from MVCCRs such that
\begin{subequations}
\label{eq:couplingMVRes}
\begin{align}
P^{\omega}_{r,t} = \tilde{P}^{\omega}_{r,t} && \forall l \in \mathcal{L}, t\in\mathcal{T}, \omega\in{\Omega}\\
Q^{\omega}_{r,t} = \tilde{Q}^{\omega}_{r,t} && \forall l \in \mathcal{L}, t\in\mathcal{T}, \omega\in{\Omega}.
\end{align}
\end{subequations}
For the LV grids connected to the MV network, power flows and voltage magnitudes seen at the respective MV nodes should be duplicated to LV's PCCs. It is given by
\begin{subequations}
\label{eq:couplingLV}
\begin{align}
P^{\omega}_{l,t} = p_{0,t}^{l,\omega} && \forall l \in \mathcal{L}, t\in\mathcal{T}, \omega\in{\Omega}\\
Q^{\omega}_{l,t} = q_{0,t}^{l,\omega} && \forall l \in \mathcal{L}, t\in\mathcal{T}, \omega\in{\Omega}\\
\eqref{eq:MV_LV_Vcouple}
%|\bar{V}^{\omega}_{l,t}| = |\bar{v}_{0,t}^{l,\omega}| && \forall l \in \mathcal{L}, t\in\mathcal{T}, \omega\in{\Omega}
\end{align}
\end{subequations}

We define augmented Lagrangian by using a sequence of Lagrangian multipliers ${y}^p_{l,t}$, ${y}^q_{l,t}$, ${y}^v_{l,t}$, ${y}^p_{r,t}$, ${y}^q_{r,t}$ for each of the coupling constraints in \eqref{eq:couplingMVRes} and \eqref{eq:couplingLV}. It is
\small
\begin{align}
&\begin{aligned}
    & L_{\rho} = \sum_{\omega \in \Omega}\sum_{t \in \mathcal{T}} \Bigg\{\left\| \bar{S}_{0,t}^{\omega} - \bar{S}^{\text{disp}}_t \right\|^2 + \sum_{r\in\mathcal{R}} f_r(P_{r,t}^{\omega}, Q_{r,t}^{\omega})\\ +
    & \sum_{l\in\mathcal{L}}\Big(\left\| \bar{s}_{0,t}^{l,\omega} - s^{l,\text{pcc}}_t \right\|^2 + \sum_{r\in\mathcal{R}_l} f^l_r(p_{r,t}^{l,\omega}, q_{r,t}^{l,\omega})\Big) +\\
    & \frac{\rho}{2}\sum_{l\in\mathcal{L}}\Big\{\left\|P_{l,t}^\omega-p_{0,t}^{l,\omega}\right\|^2 + 
    \left\|Q_{l,t}^\omega-q_{0,t}^{l,\omega}\right\|^2 + \left\||\bar{V}_{l,t}^\omega|-|\bar{v}_{0,t}^{l,\omega}|\right\|^2\Big\}  +\\ & \frac{\rho}{2}\sum_{r\in\mathcal{R}}\Big\{\left\|P^{\omega}_{r,t}-\tilde{P}_{r,t}^\omega\right\|_2^2  +
    \left\|Q^{\omega}_{r,t}-\tilde{Q}_{r,t}^\omega\right\|_2^2\Big\} + \\ 
        & \sum_{l\in\mathcal{L}}\Big\{{{y}^p_{l,t}}^T({P}_{l,t}^\omega-p_{0,t}^{l,\omega}) + {{y}^q_{l,t}}^T({Q}_{l,t}^\omega-q_{0,t}^\omega) + {{y}^v_{l,t}}^T(|\bar{V}_{l,t}^\omega|-|\bar{v}_{0,t}^{l,\omega})|\Big\} \\ & 
    \sum_{r\in\mathcal{R}}\Big\{{{y}^p_{r,t}}^T(P^{\omega}_{r,t}-\tilde{P}_{r,t}^\omega)+ {{y}^q_{r,t}}^T(Q^{\omega}_{r,t}-\tilde{Q}_{r,t}^\omega)\Big\}
    \Bigg\}.
\end{aligned}
\end{align}
\normalsize
subject to,~\eqref{eq:MVconst}, \eqref{eq:Resconst}, \eqref{eq:LVconst} and \eqref{eq:ResconstLV}.
Here, $\rho$ refers to the penalty parameter. Let define scaled dual variables $u = y/\rho$ for all the Langrange multipliers denoted by ${u}^p_{l,t}$, ${u}^q_{l,t}$, ${u}^v_{l,t}$, ${u}^p_{r,t}$, ${u}^q_{r,t}$. The above problem can be solved in following three iterative steps using the scaled-ADMM sharing problem \cite{boyd2011distributed}. The iterative steps are summarized in the \textbf{Algorithm 1}.
\normalsize

Here, $k$ refers to the iteration index of ADMM. 
First, the original variables in \eqref{eq:original_updates} are computed in parallel for each MVCCRs and downstream LV networks. The updates of the copied variables in \eqref{eq:copied_updates}, require collecting the local solutions, and it is solved by the MV aggregator. Also, the the dual variables in \eqref{eq:dual_updates} are sent to local subproblems by the aggregator. Then, the updated solutions of the copied and dual updated are disseminated to the resources. 
Eq. \eqref{eq:original_updates}, \eqref{eq:copied_updates}, and \eqref{eq:dual_updates} are solved till convergence criteria is met, i.e., when the primal and dual residual norms \cite{boyd2011distributed} reduce below a tolerance limit. For the penalty parameter $\rho$, we follow a self-adaptive approach as described in \cite{he2000alternating, boyd2011distributed}. 
%%%%%%%%%%%%%%%%%%%%%%%%%%%%%%%%%%%%%%%%%%%%%%%%
\begin{algorithm}
\caption{ADMM}\label{alg:ADMM}
\begin{algorithmic}[1]
\small
\Require $p_{0,t}^{l,\omega}(0), q_{0,t}^{l,\omega}(0), \bar{v}_{0,t}^{l,\omega}(0), \tilde{P}_{r,t}^\omega(0),
                   \tilde{Q}^\omega_{r,t}(0), P_{t}^{l,\omega}(0) $, $ Q_{t}^{l,\omega}(0), |\bar{V}_{t}^{l,\omega}(0)|,{P}_{r,t}^\omega(0),{Q}^\omega_{r,t}(0)$, $\rho>0$, $k = 0$
    \While {Convergence criteria \eqref{eq:convergence_Criteria} is not satisfied}
        \State 
        {
        Solve local variables of $l-$th LV networks ${\forall\omega\in{\Omega}, t\in\mathcal{T}}$\\
        \begin{subequations}\label{eq:original_updates}
                %\hspace{12pt} $[p_{0,t}^{l,\omega}(k+1), q_{0,t}^{l,\omega}(k+1), |\bar{v}_{0,t}^{l,\omega}(k+1)|, \forall \omega\in{\Omega}, t\in\mathcal{T}] =$
                \footnotesize
                \begin{align}
                \begin{bmatrix}
                   p_{0,t}^{l,\omega}(k+1)\\
                   q_{0,t}^{l,\omega}(k+1)\\
                   \bar{v}_{0,t}^{l,\omega}(k+1)
                \end{bmatrix} = 
                \begin{cases}
                & \begin{aligned}
                \underset{\begin{matrix}
                   \scriptstyle p_{0,t}^{l}, q_{0,t}^{l}\\
                   \scriptstyle s^{l,\text{pcc}}
                \end{matrix}}{\text{arg min}}
                 &\sum_{\omega \in \Omega}\sum_{t \in \mathcal{T}} \Bigg\{(s_{0,t}^{l,\omega} - s^{l,\text{pcc}}_t)^2+ \\
                 &\sum_{r\in\mathcal{R}_l} f^l_r(p_{r,t}^{l,\omega}, q_{r,t}^{l,\omega}) + \\ 
                  &\frac{\rho}{2}\left\|P_{l,t}^\omega(k)-p_{0,t}^{l,\omega} + u^p_{l,t}(k)\right\|^2 +\\
                  &\frac{\rho}{2}\left\|Q_{l,t}^\omega(k)-q_{0,t}^{l,\omega} + u^q_{l,t}(k)\right\|^2 +\\  & \frac{\rho}{2}\left\||\bar{V}_{l,t}^\omega(k)|-|\bar{v}_{0,t}^{l,\omega}|+ u^v_{l,t}(k)\right\|^2 \Bigg\}
                \end{aligned}\\
                & \text{subject to}:~\eqref{eq:LVconst}-\eqref{eq:ResconstLV}.
                \end{cases}
                \end{align}
                % \hspace{12pt} 
                \small
                \hspace{12pt} and, solve local variables of $r-$th MVCCRs, ${\forall\omega\in{\Omega}, t\in\mathcal{T}}$
                %$[\bar{P}_{r,t}^\omega(k+1), \bar{Q}^\omega_{r,t}(k+1), \forall \omega\in{\Omega}, t\in\mathcal{T}] = $
                \footnotesize
                \begin{align}
                \begin{bmatrix}
                   \tilde{P}_{r,t}^\omega(k+1)\\
                   \tilde{Q}^\omega_{r,t}(k+1)
                \end{bmatrix} = 
                \begin{cases}
                &\begin{aligned}
                \underset{\tilde{P}_{r,t}^\omega, \tilde{Q}_{r,t}^\omega}{\text{arg min}}
                    \sum_{\omega \in \Omega}\sum_{t \in \mathcal{T}} \Big\{f_r(\tilde{P}_{r,t}^\omega,\tilde{Q}_{r,t}^\omega) +\\
                    \frac{\rho}{2}\left\|P^{\omega}_{r,t}(k)-\tilde{P}_{r,t}^{\omega} + u^p_{r,t}(k)\right\|^2  +\\
                    \frac{\rho}{2}\left\|Q^{\omega}_{r,t}(k)-\tilde{Q}_{r,t}^{\omega} + u^q_{r,t}(k)\right\|^2\Big\}
                \end{aligned}\\
                & \text{subject to}:~\eqref{eq:Resconst}.
                \end{cases}
                \end{align}
                \end{subequations}}
        \State{ 
        \small
        Solve copied variables
        \begin{subequations}\label{eq:copied_updates}
%                $[P_{l,t}^{l,\omega}(k+1), Q_{l,t}^{l,\omega}(k+1), |\bar{V}_{l,t}^{l,\omega}(k+1)|, {P}_{r,t}^\omega(k+1),\\ {Q}^\omega_{r,t}(k+1)~\forall \omega\in{\Omega}, t\in\mathcal{T}, l\in\mathcal{L}, r\in\mathcal{R}] = $
                \footnotesize
                \begin{align}
                \begin{bmatrix}
                   P_{l,t}^{\omega}(k+1)\\
                   Q_{l,t}^{\omega}(k+1)\\
                   |\bar{V}_{l,t}^{\omega}(k+1)|\\
                   {P}_{r,t}^\omega(k+1)\\
                   {Q}^\omega_{r,t}(k+1) 
                \end{bmatrix}=
                \begin{cases}
                & \begin{aligned}
                & \underset{\begin{matrix}
                                \scriptstyle S^{\text{disp}}, P_{l,t}, \\
                                \scriptstyle Q_{l,t}, |\bar{V}_{l,t}|\\
                        \end{matrix}}{\text{arg min}} \sum_{\omega \in \Omega}\sum_{t \in \mathcal{T}} \Bigg\{(S_{0,t}^{\omega} - S^{\text{disp}}_t)^2 + \\ &
                    \sum_{l\in\mathcal{L}}\Big(\frac{\rho}{2}\left\|P_{l,t}^\omega-p_{0,t}^{l,\omega}(k) + u^p_{l,t}(k)\right\|^2 +\\&
                    \frac{\rho}{2}\left\|Q_{l,t}^\omega-q_{0,t}^{l,\omega}(k) + u^q_{l,t}(k)\right\|^2 + \\ & \frac{\rho}{2}\left\||\bar{V}_{l,t}^\omega|-|\bar{v}_{0,t}^{l,\omega}(k)|+ u^v_{l,t}(k)\right\|^2 \Big) + \\& 
                    \sum_{r\in\mathcal{R}}\Big(\frac{\rho}{2}\left\|P^{\omega}_{r,t}-\tilde{P}_{r,t}^{\omega}(k) + u^p_{r,t}(k)\right\|^2 +\\&
                    \frac{\rho}{2}\left\|Q^{\omega}_{r,t}-\tilde{Q}_{r,t}^{\omega}(k) + u^q_{r,t}(k)\right\|^2\Big)\Bigg\}
                \end{aligned}\\
                & \text{subject to}, \eqref{eq:MVconst}.
                \end{cases}
                \end{align}
                \end{subequations}}
                \footnotesize
                \small
             \State Update dual variables
             \footnotesize
                \begin{align}\label{eq:dual_updates}
                    u^p_{l,t}(k+1) &= u^p_{l,t}(k) + p_{0,t}^{l,\omega}(k+1) - P_{l,t}^\omega(k+1)~ \forall l \in\mathcal{L}\\
                    u^q_{l,t}(k+1) &= u^q_{l,t}(k) + q_{0,t}^{l,\omega}(k+1) - Q_{l,t}^\omega(k+1)~ \forall l \in\mathcal{L}\\
                    u^v_{l,t}(k+1) &= u^v_{l,t}(k) + |\bar{v}_{0,t}^{l,\omega}(k+1)|- |\bar{V}_{l,t}^\omega(k+1)|~ \forall l \in\mathcal{L}\\
                    u^p_{r,t}(k+1) &= u^p_{r,t}(k) + \tilde{P}_{r,t}^{\omega}(k+1)-P^{\omega}_{r,t}(k+1)~\forall r \in\mathcal{R}\\
                    u^q_{r,t}(k+1) &= u^q_{r,t}(k) + \tilde{Q}_{r,t}^{\omega}(k+1)-Q^{\omega}_{r,t}(k+1)~\forall r \in\mathcal{R}
                \end{align}
            \small
        \State Check convergence \eqref{eq:convergence_Criteria}.
        \State $k \gets k+1$ 
    \EndWhile
\end{algorithmic}
\end{algorithm}
%------------------------------------------%
\subsection{Convergence criteria}\label{sec:convergence_criteria}
The ADMM algorithm converges when the primal and dual residuals reduce below a feasibility tolerance bound.
The primal residual is
\small
\begin{align}
\begin{aligned}
   s_{\text{pri}}(k+1) & = \sum_{r\in\mathcal{R}}\left\|
   \begin{bmatrix}
      \tilde{P}_{r,t}^{\omega}(k+1)\\
      \tilde{Q}_{r,t}^{\omega}(k+1)
   \end{bmatrix} - 
   \begin{bmatrix}
      P^{\omega}_{r,t}(k+1)\\ 
      Q^{\omega}_{r,t}(k+1)] 
   \end{bmatrix}\right\| +  \\
   & \sum_{l\in\mathcal{L}}\left\|
   \begin{bmatrix}
      p_{0,t}^{l,\omega}(k+1)\\
      q_{0,t}^{l,\omega}(k+1)\\
      |\bar{v}_{0,t}^{l,\omega}(k+1)| 
   \end{bmatrix}-
   \begin{bmatrix}
      P_{l,t}^\omega(k+1)\\ 
      Q_{l,t}^\omega(k+1)\\
      |\bar{V}_{l,t}^\omega(k+1)|
   \end{bmatrix}
   \right\|
\end{aligned}   
\end{align}
\normalsize
and the dual residual is
\footnotesize
\begin{align}
       s_{\text{dual}}(k+1) & = \sum_{r\in\mathcal{R}}\left\|
       \begin{bmatrix}
       {P}_{r,t}^{\omega}(k+1)\\
       {Q}_{r,t}^{\omega}(k+1)
       \end{bmatrix} - 
       \begin{bmatrix}
          P^{\omega}_{r,t}(k)\\
          Q^{\omega}_{r,t}(k)  
       \end{bmatrix} \right\| + \\
   & \sum_{l\in\mathcal{L}}\left\|\begin{bmatrix}
       P_{l,t}^\omega(k+1)\\
       Q_{l,t}^\omega(k+1)\\
       |\bar{V}_{l,t}^\omega(k+1)|
   \end{bmatrix} - \begin{bmatrix}
       P_{l,t}^\omega(k)\\
       Q_{l,t}^\omega(k)\\
       |\bar{V}_{l,t}^\omega(k)|
   \end{bmatrix}\right\|.
\end{align}
\normalsize
The convergence criteria is given by
\begin{align}
    s_{\text{pri}}(k+1) \leq \epsilon_{\text{pri}}, ~\text{and}~
    s_{\text{dual}}(k+1) \leq \epsilon_{\text{dual}}\label{eq:convergence_Criteria}
\end{align}
where $\epsilon_{\text{pri}}$ and $\epsilon_{\text{dual}}$ are dynamic tolerance as defined in \cite{boyd2011distributed}.
%%%%%%%%%%%%%%%%%%%%%%%%%%%%%%%%%%%%%%%%%%%%%%%%%%%%%%%%%
\subsection{Example of controllable resource: the case of BESS} \label{sec:BESS_cost}
\subsubsection{BESS} the objective is to compute power set-points while obeying physical limits on the power rating and reservoir size. We account for BESS losses by integrating its equivalent series resistance into the network admittance matrix using the method described in \cite{stai2017dispatching}. Let the series $P_{r,t}, Q_{r,t}$ be the decision variables for active and reactive power, the BESS decision problem is the following feasibility problem:
\begin{subequations}
\label{eq:BESS_model}
\begin{align}
f_r(P_{r,t}, Q_{r,t}) = \sum_{t\in\mathcal{T}} {1}
\end{align}
The set $\Phi_r(P_{r,t}, Q_{r,t}) $ defines following set of constraints
\begin{align}
&\text{SOC}_{t} = \text{SOC}_{t-1} - P_{b,t}T_s/E^b_{\text{max}}/3600  & t\in\mathcal{T} \label{eq:SOE_update}\\
& 0 \leq (P_{b,t})^2 + (Q_{b,t})^2 \leq ({S}^b_{\text{max}})^2 & t\in\mathcal{T}   \label{eq:BETT_cap}\\
& a \le \text{SOE}_{t} \le (1-a) & t\in\mathcal{T}  \label{eq:bess_limits}
\end{align}
\end{subequations}
where, $\text{SOC}_{t}$ is the BESS state-of-charge, $T_s$ is the sampling time (900 sec in this case), ${S}^b_{\text{max}}$, and ${E}^b_{\text{max}}$ are the power and reservoir capacities respectively, 
and $0 \leq a < 0.5$ is a fixed parameter to specify a margin on SOE limits. % for hard constraints on the BESS energy budget. 
%Eq. \eqref{eq:SOE_update} expresses the state of energy. 
The constraint \eqref{eq:BETT_cap} is to restrict the battery's apparent power within its four-quadrant converter capability.
\section{Simulation Results and discussions} 
\subsection{Multi-grid test case}
\begin{figure*}[t]
    \centering
    \includegraphics[width=0.9\linewidth]{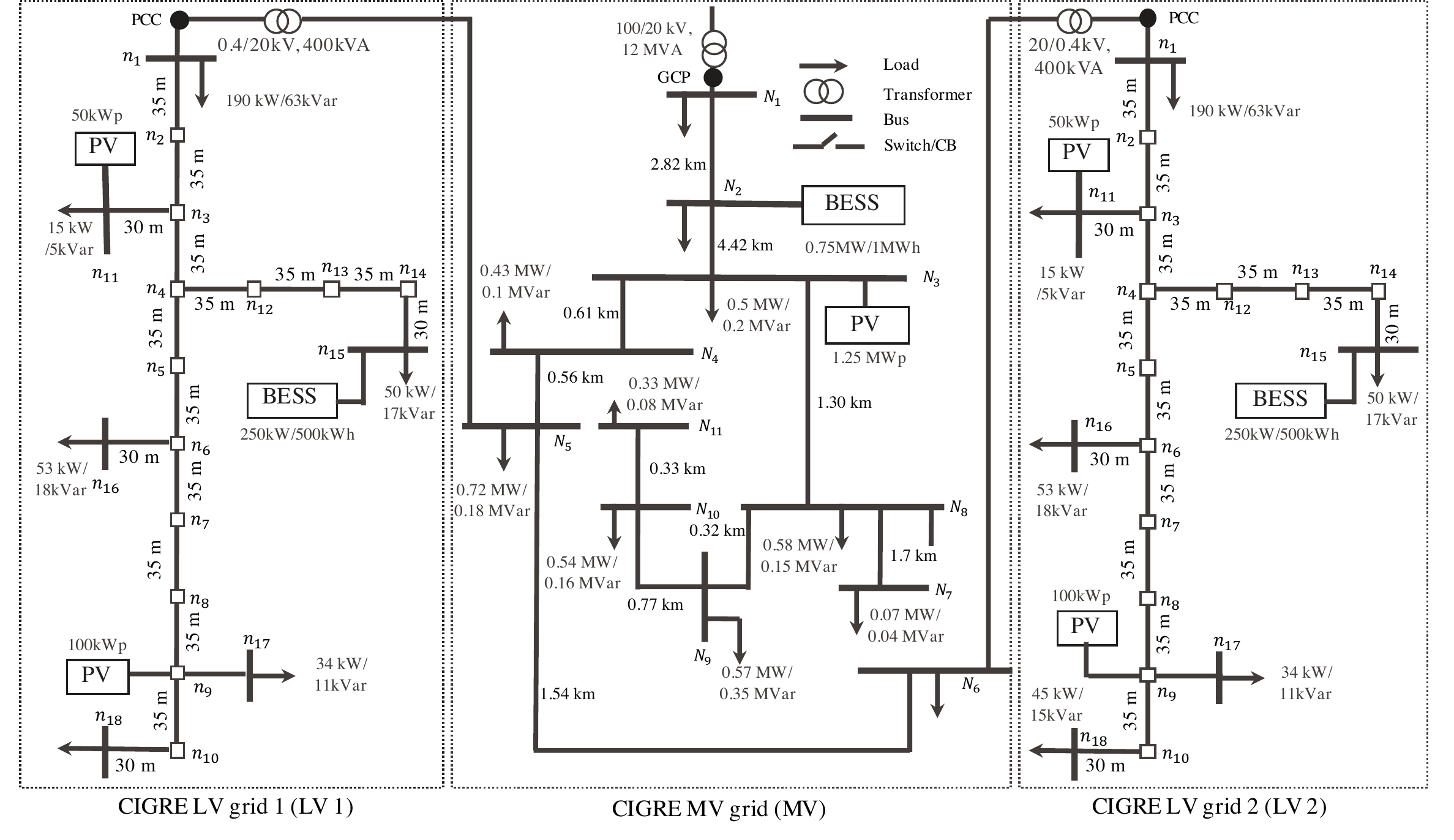}
    \caption{{Multi-grid test case: CIGRE MV and benchmark LV networks}}
    \label{fig:multigrid}
\end{figure*}
\begin{figure*}[h!]
\centering
\subfloat[]{\includegraphics[width=0.45\linewidth]{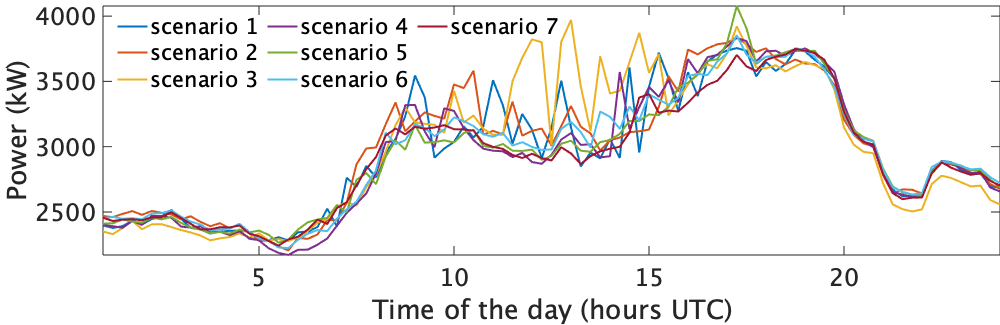}
\label{fig:PMV_sc}}
\subfloat[]{\includegraphics[width=0.45\linewidth]{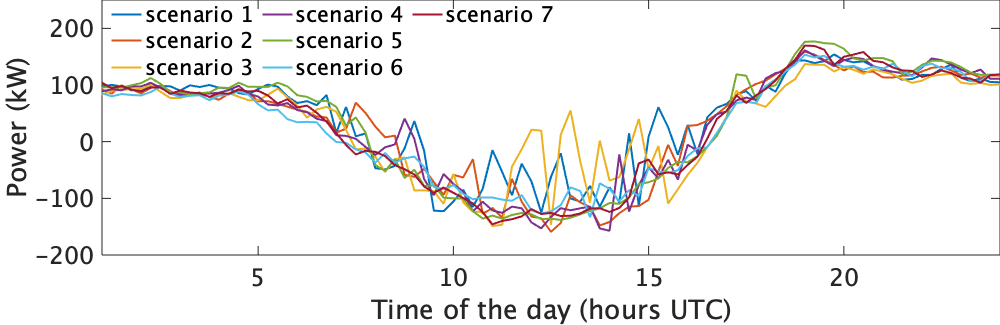}
\label{fig:PLV_sc}}\\
\subfloat[]{\includegraphics[width=0.45\linewidth]{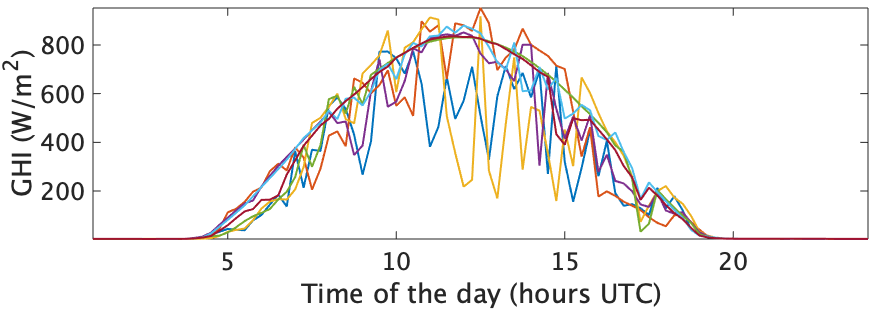}
\label{fig:GHI_sc}}
\subfloat[]{\includegraphics[width=0.45\linewidth]{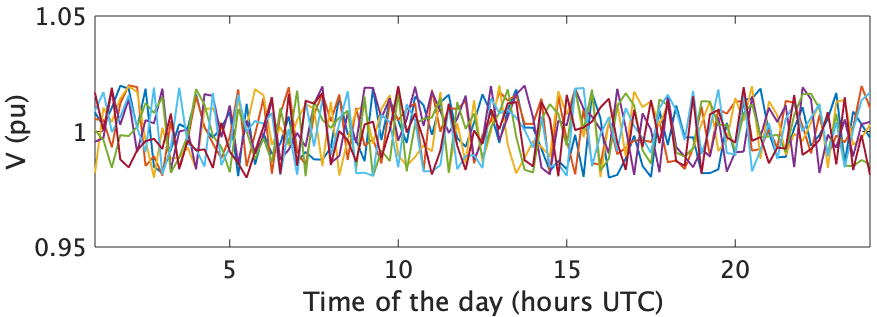}
\label{fig:V_slack_sc}}
\caption{Day-ahead scenarios of aggregated nodal active powers (in kW) for (a) MV network and (b) LV1 network, (c) GHI (in W/m$^2$) and (d) imposed voltage (in pu) at MV's GCP.} \label{fig:PQVscenarios}
\end{figure*}
%
%
%\clearpage
\begin{figure}[!h]
\centering
\subfloat[Computed dispatch plan and aggregated active power at the MV's GCP.]{\includegraphics[width=0.95\linewidth]{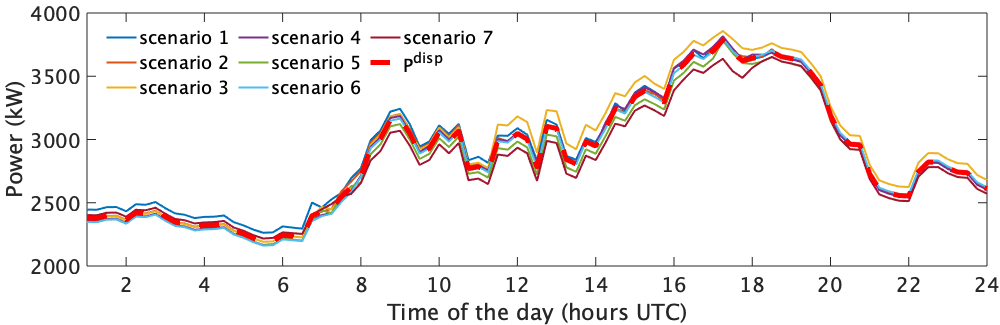}
\label{fig:MVslack_unc}}\\
\subfloat[Active power injection provided by MV's BESS for different scenarios.]{\includegraphics[width=0.95\linewidth]{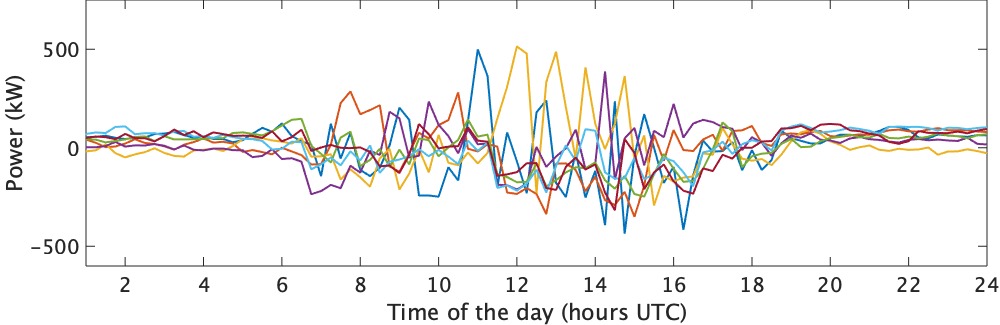}
\label{fig:BESSPMV_unc}}\\
\subfloat[SOC of MV's BESS for different scenarios.]{\includegraphics[width=0.95\linewidth]{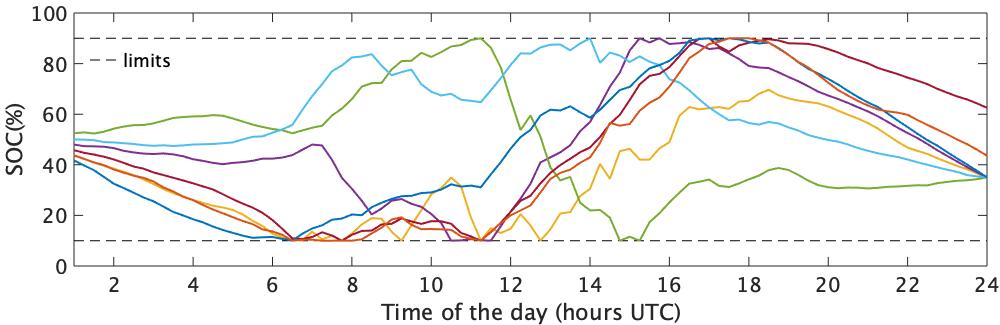}
\label{fig:BESSSOCMV_unc}}\\
\subfloat[Computed dispatch plan and aggregated active power at the LV1's GCP.]{\includegraphics[width=0.95\linewidth]{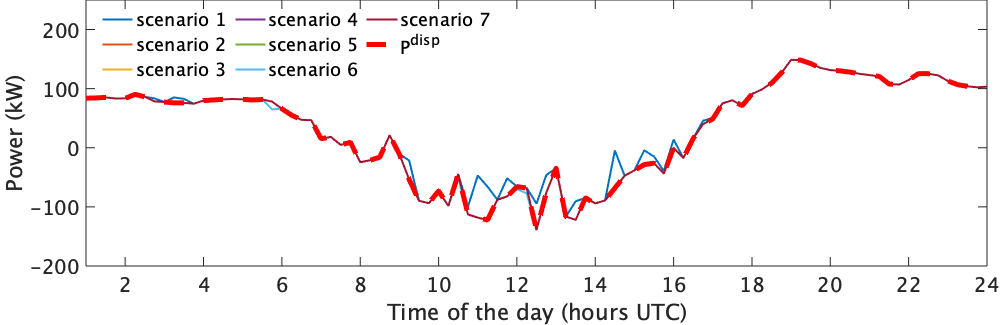}
\label{fig:LVslack_unc}}\\
\subfloat[Active power injection provided by LV1's BESS for different scenarios.]{\includegraphics[width=0.95\linewidth]{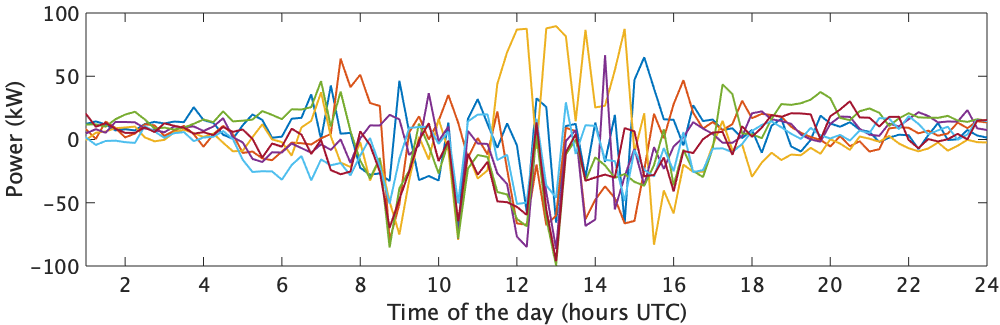}
\label{fig:BESSPLV_unc}}\\
\subfloat[SOC of LV1's BESS for different scenarios.]{\includegraphics[width=0.95\linewidth]{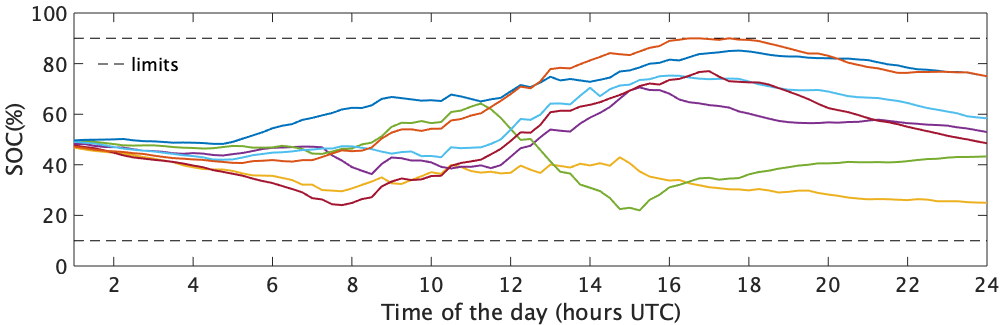}}
\label{fig:BESSSOCLV_unc}
\caption{Dispatch plan computation for no-coordination case among MV and LV networks.}  
\label{fig:uncoordinated}
\end{figure}
\begin{figure}[!h]
\centering
\subfloat[Computed dispatch plan and aggregated active power at the MV's GCP.]{\includegraphics[width=0.95\linewidth]{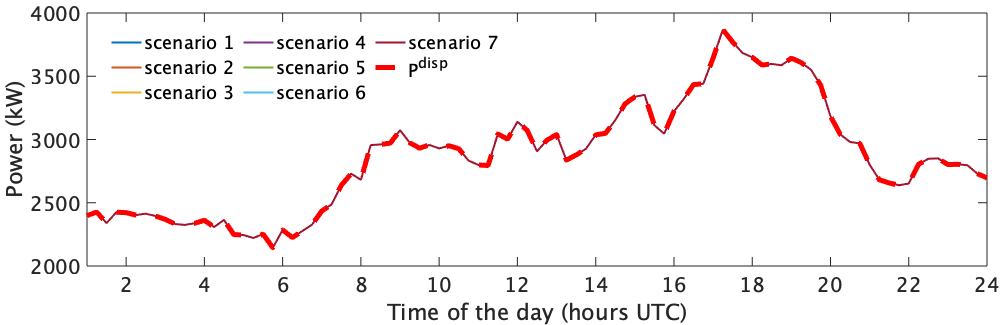}
\label{fig:MVslack_cor}}\\
\subfloat[Active power injection provided by MV's BESS for different scenarios.]{\includegraphics[width=0.95\linewidth]{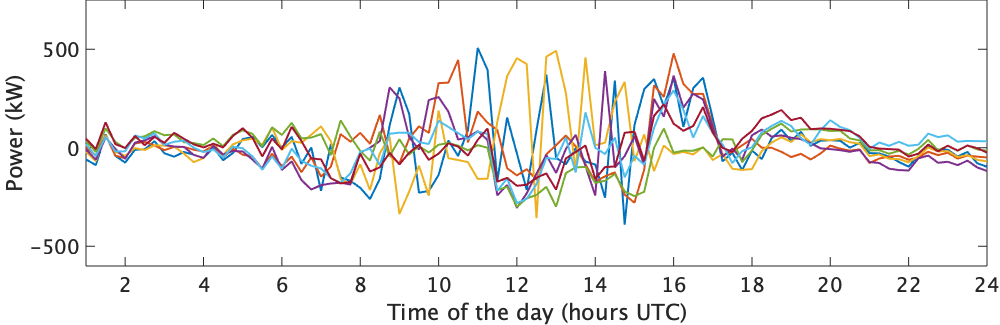}
\label{fig:BESSPMV_cor}}\\
\subfloat[SOC of MV's BESS for different scenarios.]{\includegraphics[width=0.95\linewidth]{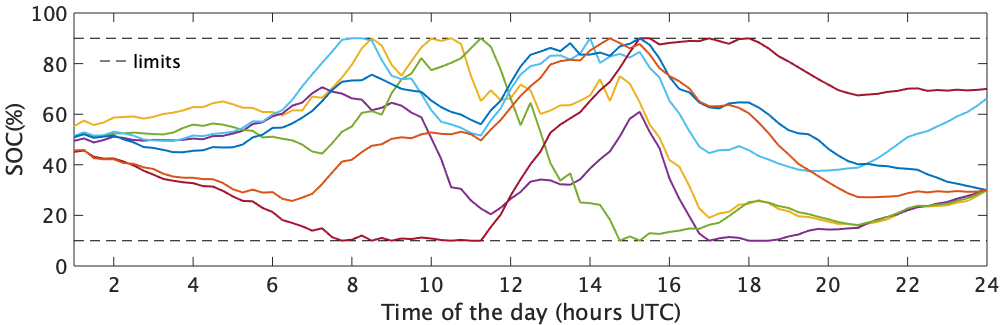}
\label{fig:BESSSOCMV_cor}}\\
\subfloat[Computed dispatch plan and aggregated active power at the LV1's GCP.]{\includegraphics[width=0.95\linewidth]{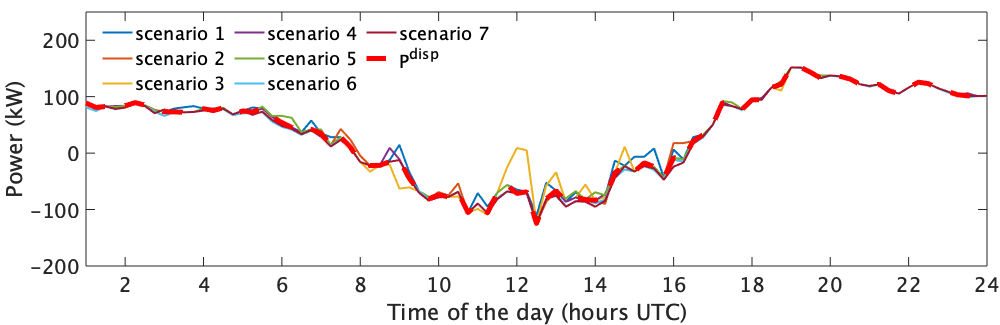}
\label{fig:LVslack_cor}}\\
\subfloat[Active power injection provided by LV1's BESS for different scenarios.]{\includegraphics[width=0.95\linewidth]{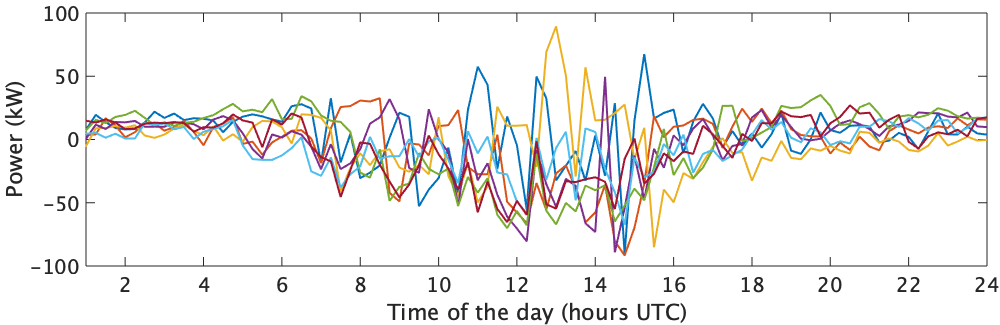}
\label{fig:BESSPLV_cor}}\\
\subfloat[SOC of LV1's BESS for different scenarios.]{\includegraphics[width=\linewidth]{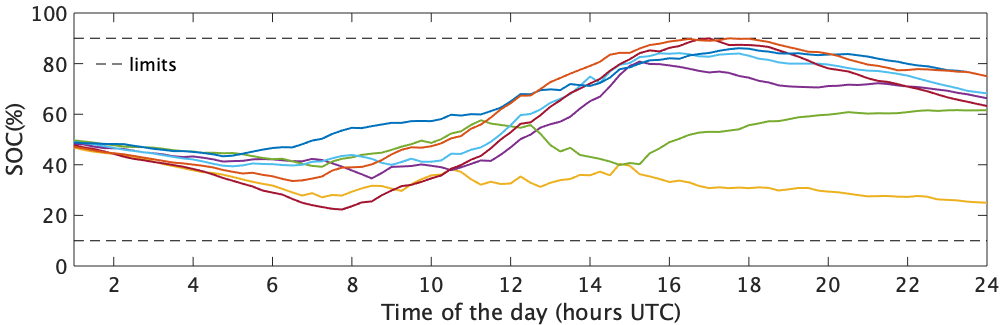}}
\caption{Dispatch plan computation using the proposed multi-grid dispatch by ADMM-based coordination of the MV and LV systems.}  \label{fig:coordinated}
\end{figure}
We validate the proposed day-ahead dispatch computation for a multi-grid system shown in Fig.~\ref{fig:multigrid}. It consists of two identical LV systems connected to nodes $N_5$ and $N_6$ of the MV network. The MV and LV networks are CIGRE benchmark test-cases \cite{CIGREREF} of nominal power/voltage ratings of 12~MVA/20kV and 400~kVA/400~V respectively. The nominal demands and generation units are shown in the Fig.~\ref{fig:multigrid}. It also shows the capacity of the PV generation units and controllable resources. The capacity and the sites of the controllable and PV units are also summarized in Table~\ref{tab:DERs}.

\begin{table}[h!]
    \centering
    \caption{Sites and sizes of BESS and PV units}
    \begin{tabular}{|c|c|c|c|c|}
        \hline
        &   \multicolumn{2}{|c|}{\bf{PV}} & \multicolumn{2}{|c|}{\bf{BESS}}\\
        \hline
                 & node & size   & node & size\\
        \hline
        \bf{MV} &   $N_3$  & 1.25 [MWp]  &  $N_2$   & 0.75 [MW]/1.0 [MWh]   \\
        \hline
        \bf{LV1} &   $n_9$  &    100 [kWp] &  $n_{15}$    & 250 [kW]/500 [kWh]   \\
                &   $n_{11}$  &    50 [kWp] &     &  \\
        \hline
        \bf{LV2} &   $n_9$  &    100 [kWp] &  $n_{15}$    & 250 [kW]/500 [kWh]   \\
                &   $n_{11}$  &    50 [kWp] &     &  \\
        \hline
    \end{tabular}
    \label{tab:DERs}
\end{table}
\subsection{Day-ahead scenarios}
Since the proposed framework is a scenario-based stochastic optimization, we forecast the uncertainties of the demand and generation by a set of scenarios that are forecasted in day-ahead. The scenarios are modeled using the historical data, we use the scenario reduction and forecasting strategy from \cite{gupta2020grid}. For the load, it selects $N_{\Omega}$ scenarios\footnote{The selection of ideal $N_\Omega$ is beyond the scope of this work, we determine $N_\Omega = 7$ that covers 95 percentile of the variation during the realization.} of 1-day time series of historical measurements of demand according to the day-type (working day, weekend, day of the weak, the period of the year).
The PV generation is forecasted starting from the predictions of the time-series of the global horizontal irradiance (GHI). We use GHI predictions scheme of \cite{gupta2020grid}. We also model the uncertainty on the voltage imposed at the MV network's GCP from the upstream transmission system. This is done by obtaining the cumulative distribution function of the voltage variations using historical data from a real Swiss MV distribution network. As per the historical data, the voltage at the GCP was found to be varying uniformly within [0.98, 1.02]~pu, so we generated the scenarios by a uniform distribution. Fig~\ref{fig:PMV_sc} and~\ref{fig:PLV_sc} shows the scenarios for aggregated prosumption at MV's GCP and LV1's PCC respectively. Fig~\ref{fig:GHI_sc} shows the scenarios for GHI, they are same for both LV and MV systems. Fig~\ref{fig:V_slack_sc} shows the profile for MV's GCP voltage in per units.

\subsection{Simulation results}
Using the day-ahead forecast scenarios of the load, generation and MV GCP voltage, we compute the day-ahead dispatch plan for the multi-grid system of Fig.~\ref{fig:multigrid}. For the sake of brevity, we show the computed dispatch plan at the MV system GCP and at the PCC of one of the LV network (as both LV networks are identical). We simulate two cases: 
\begin{itemize}
    \item \textit{No coordination:}  we define a base case where the MV system does not account for the flexibility of downstream LV networks. In this case, MV and LV networks operate as standalone systems, i.e. LV networks compute their dispatch plan and send it to the MV network. Then, MV system computes its dispatch plan by modeling LV networks as uncontrollable load.\\
    \item \textit{Multi-grid dispatch using ADMM-based coordination:} we solve the proposed distributed dispatch computation of \textbf{Algorithm 1}. As developed in Sec.~\ref{sec:ADMM_dispatch}, this scheme coordinates with the downstream LV networks while computing the dispatch plans.
\end{itemize}
\subsubsection{No coordination among MV and LV systems}
Figure~\ref{fig:uncoordinated} presents the simulation results for dispatch plan computation for no coordination case. Figure~\ref{fig:MVslack_unc}, \ref{fig:LVslack_unc} shows the computed dispatch plan and the compressed scenarios of the power at the GCP/PCC for MV and LV1 networks respectively. Figure~\ref{fig:BESSPMV_unc}, ~\ref{fig:BESSSOCMV_unc} and Fig~\ref{fig:BESSPLV_unc}, ~\ref{fig:uncoordinated}f shows the BESS active power injections and SOC for MV and LV1 networks.
As it can be observed, the computed dispatch plan for the MV network is not tracked in all the day-ahead scenarios. This happens because the BESS (only controllable resource) reaches its state-of-charge bounds for those day-ahead scenarios.
\subsubsection{Multi-grid dispatch using ADMM-based coordination among MV and LV systems}
Figure~\ref{fig:coordinated} shows the dispatch plan computation when ADMM-based coordination was used. As it can be observed, the MV system manage to respect the dispatch plan in all the day-ahead scenarios thanks to the contribution from the LV systems at node $N_5$ and $N_6$ as well as the BESS at node $N_3$. Although the dispatch plan in case of LV network is slightly worse than one without coordination, it helps the MV network to respect the dispatch plan for all day-ahead scenarios.
To quantify the difference in the tracking performance, Table~\ref{tab:Performance} shows the performance comparison of two cases in terms of following metrics. 
The metrics are maximum absolute error (MAE)
\begin{align}
    \text{MAE} = {\underset{\omega\in\Omega, t\in\mathcal{T}}{\text{max}}|P^{\text{disp}}_t - P_{0,t}^{\omega}|}%{\underset{\omega\in\Omega, t\in\mathcal{T}}{\text{max}}|P^{\text{disp}}_t|}\times100,
\end{align}
and normalised mean of sum of absolute differences (SAD)
\begin{align}
    \text{NSAD} = {\underset{\omega\in\Omega, t\in\mathcal{T}}{\text{sum}}|P^{\text{disp}}_t - P_{0,t}^{\omega}|}/{\underset{\omega\in\Omega, t\in\mathcal{T}}{\text{sum}}|P^{\text{disp}}_t|}\times100.
\end{align}
The comparison in Table~\ref{tab:Performance} shows that the MAE and NSAD incurred on the dispatch reduced from 147~kW and 8.5\% to 1.5~kW and 1e-2~\% respectively using the proposed ADMM-coordinated multi-dispatch algorithm. Although LV networks perform poorly in terms of NSAD, the MAE improves. This is because LV networks are providing flexibility to the upstream MV network to track its dispatch with high fidelity.
%%%%%%%%%%%%%%%%%%%%%%%%%%%%%%%%%%%%%%%%%%%%%%%%%%%%%%%%%%%
%%%%%%%%%%%%%%%%%%%%%%%%%%%%%%%%%%%%%%%%%%%%%%%%%%%%%%%%
\begin{table}[h!]
    \centering
    \caption{Dispatch performance}
    \begin{tabular}{|c|c|c|c|c|}
        \hline
        &   \multicolumn{2}{|c|}{\bf{No coordination}} & \multicolumn{2}{|c|}{\bf{Proposed multi-grid dispatch using}}\\
        &   \multicolumn{2}{|c|}{\bf{}} & \multicolumn{2}{|c|}{\bf{ADMM-based coordination}}\\

        \hline
                 & MAE(kW) & NSAD(\%)   & MAE(kW) & NSAD(\%)\\
        \hline
        \bf{MV} &  147   & 8.5 &  1.5  & 1e-2   \\
        \hline
        \bf{LV1} &  71.2  & 6.4 &  59  & 8.5   \\
        \hline
        \bf{LV2} &  71.2  & 6.4 &  59   & 8.5   \\
        \hline
    \end{tabular}
    \label{tab:Performance}
\end{table}
\section{Conclusion}
This work developed a framework to compute aggregated day-ahead dispatch plans of multiple and interconnected distribution grids operating at different voltage levels. It is achieved by accounting for the flexibility as well as the uncertainty of the downstream networks. The problem was formulated to determine the day-ahead dispatch plan of an MV network accounting for the flexibility offered by the downstream LV networks and from other MV-connected controllable resources.
The problem was formulated as scenario-based stochastic optimization where the day-ahead forecasts provide the uncertain load and generation scenarios. The optimization problem was solved by an ADMM-based distributed optimizations scheme guaranteeing better scalability and inter-grid-layer privacy.

The proposed framework was validated by simulating the CIGRE MV network connected with two identical CIGRE LV systems, controllable resources such as BESS, and stochastic resources such as PV generation units. The simulation concluded that the MV network manages to cover all the stochastic scenarios when downstream LV networks coordinate in providing flexibility to the MV network. In contrast, the MV aggregator failed to satisfy all the stochastic scenarios in no coordination case.

Future work would experimentally validate this framework on an actual interconnected distribution grid of the EPFL campus consisting of an MV network interfaced with an LV network.
\appendix
\subsection{Relaxation of the non-convex power factor constraint}
\label{sec:convexification}
We introduce two variables $P_{0,t}^{+, \omega}$ and $P_{0,t}^{-, \omega}$ such that
\begin{align}
P_{0,t}^{\omega} = P_{0,t}^{+, \omega} - P_{0,t}^{-, \omega} \label{eq:relaxation}
\end{align}
and replace Eq.~\eqref{eq:Q_cons_disp} with the following set of linear constraints:
\small
\begin{align}
    &{P_{0,t}^{+,\omega}} + {P_{0,t}^{-,\omega}} \ge Q_{0,t}^{\omega}\tan(\pi/2 -\theta_m) \label{eq:pf1}\\
    &{P_{0,t}^{+,\omega}} + {P_{0,t}^{-,\omega}} \ge -Q_{0,t}^{\omega}\tan(\pi/2 -\theta_m) \label{eq:pf2}\\
    &{P_{0,t}^{+,\omega}} \ge 0, {P_{0,t}^{-,\omega}} \ge 0\label{eq:pf4},
\end{align}
\normalsize
where $\theta_m$ refers to the angle corresponding to $\cos(\theta)_{\text{min}}$. The two terms of \eqref{eq:relaxation} ($p_{0,t}^{+, \omega},p_{0,t}^{-, \omega}$) should be mutually exclusive. To this end, we augment the cost function \eqref{eq:costdispatch} with the following new term 
\begin{align}
\sum_{\omega \in \Omega}\sum_{t \in \mathcal{T}}\nu\big(({P_{0,t}^{+,\omega}})^2 +({P_{0,t}^{-,\omega}})^2\big)
\end{align}
that promotes $P_{0,t}^{+, \omega},P_{0,t}^{-, \omega}$ being mutually exclusive, where $v\geq0$ weighs the significance of obeying power factor constraints.
Same procedure is used for LV systems.
\bibliographystyle{IEEEtran}
\bibliography{biblio.bib}
\end{document}